\documentclass[aps,prb,twocolumn,showpacs,floatfix]{revtex4}

\usepackage{graphicx}

\begin{document}

\title{Fluctuating charge density waves in the Hubbard model}

\author{Alexei Sherman}
 \affiliation{Institute of Physics, University of Tartu, Riia 142,
 51014 Tartu, Estonia}
\author{Michael Schreiber}
 \affiliation{Institut f\"ur Physik, Technische Universit\"at, 09107
 Chemnitz, Germany}

\date{\today}

\begin{abstract}
The charge susceptibility of the two-dimensional repulsive Hubbard model is investigated using the diagram technique developed for the case of strong correlations. In this technique, a power series in the hopping constant is used. It is shown that once the Fermi level crosses one of the Hubbard subbands a sharp peak appears in the momentum dependence of the static susceptibility. With further departure from half-filling the peak transforms to a ridge around the $\Gamma$ point. In the considered range $0\leq|1-\bar{n}|\alt 0.2$ of the electron filling $\bar{n}$ the static susceptibility is finite which points to the absence of the long-range charge ordering. However, for $|1-\bar{n}|\approx 0.12$ the susceptibility maxima are located halfway between the center and the boundaries of the Brillouin zone. In this case an interaction of carriers with tetragonal distortions can stabilize the charge density wave with the wavelength of four lattice spacings, as observed experimentally in the low-temperature tetragonal phase of lanthanum cuprates. In the range of parameters inherent in cuprate perovskites the character of the susceptibility evolution with $\bar{n}$ depends only weakly on the ratio of the nearest-neighbor hopping constant to the Hubbard repulsion and on details of the initial band structure. The location of the susceptibility maxima in the Brillouin zone is mainly determined by the value of $\bar{n}$.
\end{abstract}


\maketitle

\section{Introduction}
A variety of numerical methods have been used to elucidate the existence  of phase separation in the ground states of the Hubbard and the related $t$-$J$ models. The states with charge density waves (CDW, stripes) were obtained in Refs.~\onlinecite{Hizhnyakov,Zaanen,Machida,Giamarchi,An,White} using the mean-field approximation, the variational principle with the Gutzwiller-type variational functions and the density matrix renormalization group calculations. However, results of Monte Carlo simulations \cite{Vilk,Sorella,Anisimov} and cluster calculations \cite{Maier,Aichhorn} have cast doubt on this finding. Hence different techniques give different results which reflects the fact that nearly degenerate low-lying states of the models have different nature.

Experimentally static stripes were observed \cite{Tranquada95,Tranquada96,Fujita,Kimura} in the low-temperature tetragonal (LTT) phase of lanthanum cuprates La$_{1.6-x}$Nd$_{0.4}$Sr$_x$CuO$_4$ and La$_{2-x}$Ba$_x$CuO$_4$.
One of the manifestations of the stripe formation is the anomalous suppression of superconductivity near the hole concentration $x=\frac{1}{8}$ in La$_{2-x}$Ba$_x$CuO$_4$. A weaker suppression of $T_c$ near this hole concentration is also observed in La$_{2-x}$Sr$_x$CuO$_4$ in the low-temperature orthorhombic (LTO) phase.\cite{Takagi} The mentioned phases are characterized by tilts of the CuO$_6$ octahedra about axes (LTT) and diagonals (LTO) of the Cu-O planes. These experimental observations suggest that the interaction of carriers with respective phonons plays an essential role in the stripe stabilization.

For investigating instabilities of a system it is convenient to use corresponding static susceptibilities. Similar to the divergence of the magnetic susceptibility, which points to the establishment of the long-range magnetic order while finite maxima indicate short-range ordering, a divergence of the charge susceptibility means the appearance of the CDW, while finite peaks are manifestations of respective charge fluctuations. In this paper we calculate the charge susceptibility of the two-dimensional repulsive Hubbard model using the strong-coupling diagram technique. In this technique,\cite{Vladimir,Metzner,Pairault,Sherman06} on the assumption of strong electron correlations Green's functions are calculated using the expansion in powers of the hopping constant. The terms of this expansion are expressed by means of site cumulants of electron creation and annihilation operators.

It is found that once the Fermi level crosses one of the Hubbard subbands a sharp maximum appears in the momentum dependence of the static susceptibility near the $\Gamma$ point. With further departure of the electron filling $\bar{n}$ from the half-filling value $\bar{n}=1$ the maximum transforms to a ridge around the $\Gamma$ point. In the considered range of the electron filling $0\leq|1-\bar{n}|\alt 0.2$  the static susceptibility is finite which implies the absence of the long-range charge ordering. However, obtained results give some insight into the way in which phonons can stabilize stripes. For $|1-\bar{n}|\approx 0.12$ the susceptibility maxima are located halfway between the center and the boundaries of the Brillouin zone. In this case an interaction of carriers with the tetragonal distortions gives a maximal energy gain for the CDW with the wavelength of four lattice spacings, as observed experimentally in the LTT phase of lanthanum cuprates. The calculations were carried out for the $t$-$t'$-$U$ Hubbard model for the ratios of the Hubbard repulsion to the nearest-neighbor hopping constant $U/|t|=8$ and 12, and for the next-nearest-neighbor hopping constant $t'=0$ and $-0.3t$. The above-discussed evolution of the susceptibility with $\bar{n}$ is not changed qualitatively with the variation of $U/t$ and $t'/t$. The shape of the susceptibility and the location of its maxima in the Brillouin zone are mainly determined by the value of $\bar{n}$. The mentioned values of the parameters belong to the parameter range which is widely believed to be suitable for cuprate perovskites.

Main formulas used in the calculations are given in the following
section. The discussion of the obtained results and their comparison
with Monte Carlo simulations are carried out in Sec.~III\@.
Concluding remarks are presented in Sec.~IV.

\section{Main formulas}
The Hubbard model is described by the Hamiltonian
\begin{equation}\label{Hamiltonian}
H=\sum_{\bf ll'\sigma}t_{\bf ll'}a^\dagger_{\bf l\sigma}a_{\bf
 l'\sigma}+\frac{U}{2}\sum_{\bf l\sigma}n_{\bf l\sigma}n_{\bf
 l,-\sigma},
\end{equation}
where $a^\dagger_{\bf l\sigma}$ and $a_{\bf l\sigma}$ are the electron
creation and annihilation operators, ${\bf l}$ labels sites of the
square plane lattice, $\sigma=\uparrow$ or $\downarrow$ is the spin projection, $t_{\bf ll'}$ and $U$ are hopping and on-site repulsion constants, and $n_{\bf l\sigma}=a^\dagger_{\bf l\sigma}a_{\bf l\sigma}$. Below we consider the case where only the nearest-neighbor $t$ and next-nearest-neighbor $t'$ hopping constants are nonzero.

As mentioned above, in the considered case of strong correlations, $U\gg|t|,\,t'$, we use the strong-coupling diagram technique \cite{Vladimir,Metzner,Pairault,Sherman06} for calculating the charge
Green's function
\begin{equation}\label{Green}
B_{\sigma'\sigma}({\bf l'\tau',l\tau})=\langle{\cal T}\delta n_{\bf l'\sigma'}(\tau')\delta n_{\bf l\sigma}(\tau)\rangle,
\end{equation}
where $\delta n_{\bf l\sigma}=n_{\bf l\sigma}-\langle n_{\bf l\sigma} \rangle$ is the deviation of the electron occupation number from its mean value, the angular brackets denote the statistical averaging with the Hamiltonian $${\cal H}=H-\mu\sum_{\bf l\sigma}n_{\bf
l\sigma},$$ $\mu$ is the chemical potential, ${\cal T}$ is the
time-ordering operator which arranges other operators from right to
left in ascending order of times $\tau$. The time evolution of operators in~(\ref{Green}) is also determined by the Hamiltonian ${\cal H}$,
$$O(\tau)=\exp({\cal H}\tau)O\exp(-{\cal H}\tau).$$

Using the strong-coupling diagram technique one can convince oneself that equations for $B$ are similar to those derived for the spin Green's function in Ref.~\onlinecite{Sherman07}. After the Fourier transformation these equations read
\begin{eqnarray}
B_{\sigma'\sigma}(q)&=&-\delta_{\sigma'\sigma}\frac{T}{N}\sum_{p_1}
 G(p_1)G(q+p_1)\nonumber\\
&+&\left(\frac{T}{N}\right)^2\sum_{p_1p_2}\Pi(p_1)\Pi(p_2)\Pi(q+p_1)
 \Pi(q+p_2) \nonumber\\
&\times&\Lambda_{\sigma'\sigma}(p_1,q+p_1,q+p_2,p_2), \label{GF}
\end{eqnarray}
\begin{eqnarray}
&&\Lambda_{\sigma'\sigma}(p_1,q+p_1,q+p_2,p_2)\nonumber\\
&&\quad=\lambda_{\sigma'\sigma}(p_1,q+p_1,q+p_2,p_2)\nonumber\\
&&\quad-\frac{T}{N}\sum_{p_3\sigma_1}
 \lambda_{\sigma'\sigma_1}(p_1,q+p_1,q+p_3,p_3)\Theta(p_3)
 \Theta(q+p_3)\nonumber\\
&&\quad\times\Lambda_{\sigma_1\sigma}(p_3,q+p_3,q+p_2,p_2). \label{Bethe}
\end{eqnarray}
Here the combined indices $q=({\bf k},i\omega_\nu)$ and $p_j=({\bf
k}_j,i\omega_{n_j})$ were introduced, $\omega_\nu=2\nu\pi T$ and $\omega_n=(2n+1)\pi T$ are the boson and fermion Matsubara frequencies with the temperature $T$, ${\bf k}$ is the wave vector, $G(p)=\langle\langle a_{\bf k\sigma}|a^\dagger_{\bf k\sigma}\rangle\rangle$ is the electron Green's function, $\Pi(p)=1+t_{\bf k}G(p)$, $t_{\bf k}$ is the Fourier transform of the hopping constants which is equal to $t_{\bf k}=2t[\cos(k_x)+\cos(k_y)] +4t'\cos(k_x)\cos(k_y)$ in the considered case, the lattice spacing is taken as the unit of length, $\Theta(p)=t_{\bf k} \Pi(p)$ is the renormalized hopping, $\Lambda_{\sigma'\sigma}(p_1,p+p_1,p+p_2,p_2)$ is the sum of
all four-leg diagrams, $\lambda_{\sigma'\sigma}(p_1,p+p_1,p+p_2,p_2)$ is its irreducible subset, and $N$ is the number of sites.

The main difference between Eqs.~(\ref{GF}), (\ref{Bethe}) and the respective equations for the spin Green's function\cite{Sherman07} is in the spin indices of the irreducible four-leg diagrams. For the transversal spin Green's function considered in Ref.~\onlinecite{Sherman07} only $\lambda_{\uparrow\downarrow}(p_1,p+p_1,p+p_2,p_2)$ enters into the equations.

Equations~(\ref{GF}) and~(\ref{Bethe}) can be somewhat simplified if we take into account the invariance of Hamiltonian~(\ref{Hamiltonian}) with respect to rotations of the spin quantization axis.\cite{Fradkin} This invariance leads to the following symmetry relations: $$B_{\uparrow\uparrow}(q)=B_{\downarrow\downarrow}(q),\quad B_{\downarrow\uparrow}(q)=B_{\uparrow\downarrow}(q)$$ and analogously for $\Lambda_{\sigma'\sigma}$ and $\lambda_{\sigma'\sigma}$. Using these relations we find from Eqs.~(\ref{GF}) and~(\ref{Bethe})
\begin{eqnarray}
B(q)&=&-\frac{T}{N}\sum_{p_1}G(p_1)G(q+p_1)\nonumber\\
&+&\left(\frac{T}{N}\right)^2\sum_{p_1p_2}\Pi(p_1)\Pi(p_2)\Pi(q+p_1)
 \Pi(q+p_2) \nonumber\\
&\times&\Lambda(p_1,q+p_1,q+p_2,p_2), \label{GF2}
\end{eqnarray}
\begin{eqnarray}
&&\Lambda(p_1,q+p_1,q+p_2,p_2)=\lambda(p_1,q+p_1,q+p_2,p_2)\nonumber\\
&&\quad-\frac{T}{N}\sum_{p_3}\lambda(p_1,q+p_1,q+p_3,p_3)\Theta(p_3)
 \Theta(q+p_3)\nonumber\\
&&\quad\times\Lambda(p_3,q+p_3,q+p_2,p_2), \label{Bethe2}
\end{eqnarray}
where
\begin{eqnarray}
&&B(q)=\frac{1}{2}\sum_{\sigma'\sigma}B_{\sigma'\sigma}(q),\nonumber\\
&&\Lambda(p_1,q+p_1,q+p_2,p_2)\nonumber\\
&&\quad=\frac{1}{2}\sum_{\sigma'\sigma} \Lambda_{\sigma'\sigma}(p_1,q+p_1,q+p_2,p_2)\label{newvar}\\
&&\lambda(p_1,q+p_1,q+p_2,p_2)\nonumber\\
&&\quad=\frac{1}{2}\sum_{\sigma'\sigma} \lambda_{\sigma'\sigma}(p_1,q+p_1,q+p_2,p_2).\nonumber
\end{eqnarray}

\begin{figure}[t]
\centerline{\includegraphics[width=6cm]{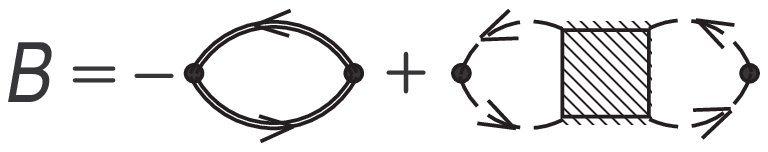}}\vspace{4ex}
\centerline{\includegraphics[width=6cm]{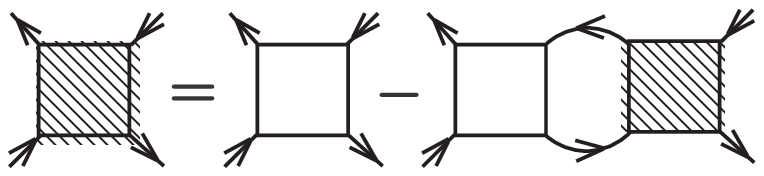}}
\caption{Diagrams corresponding to Eqs.~(\protect\ref{GF2}) and~(\protect\ref{Bethe2}).} \label{Fig1}
\end{figure}

Diagrams corresponding to Eqs.~(\ref{GF2}) and~(\ref{Bethe2}) are plotted in Fig.~\ref{Fig1}.
In these diagrams, the dual line indicates the electron Green's function $G(p)$, the shaded and open squares stand for $\Lambda$ and $\lambda$, respectively, the single directed line between the squares is the renormalized hopping $\Theta(p)$, and the dashed line is the external line of the four-leg diagram with the inserted irreducible two-leg diagrams. The dashed line corresponds to $\Pi(p)$.

In the following consideration, we simplify the general equations~(\ref{GF2}) and~(\ref{Bethe2}) by neglecting the irreducible two-leg diagrams in the external and internal lines of the four-leg diagrams. In this approximation quantities $\Pi(p)$ and $\Theta(p)$ in Eqs.~(\ref{GF2}) and~(\ref{Bethe2}) are substituted by 1 and $t_{\bf k}$, respectively. Besides, we use the lowest-order irreducible four-leg diagram instead of  $\lambda_{\sigma'\sigma}(p_1,q+p_1,q+p_2,p_2)$. This four-leg diagram is described by the second-order cumulant of electron operators
\begin{eqnarray}
K_2^{\sigma'\sigma}(\tau',\tau,\tau'_1,\tau_1)&=&\langle
 {\cal T}\bar{a}_{\sigma'}(\tau')a_{\sigma'}(\tau) \bar{a}_{\sigma}(\tau'_1)a_{\sigma}(\tau_1)\rangle_0\nonumber\\
&&-K_1(\tau',\tau)K_1(\tau'_1,\tau_1)\nonumber\\
&&+K_1(\tau',\tau_1)K_1(\tau'_1,\tau)\delta_{\sigma'\sigma},
 \label{cumulant}
\end{eqnarray}
where the subscript 0 of the angular brackets indicates that the averaging and time dependencies of operators are determined by the site Hamiltonian
\begin{equation}\label{siteH}
H_{\bf l}=\sum_\sigma[(U/2)n_{\bf l\sigma}n_{\bf l,-\sigma}-\mu
n_{\bf l\sigma}],
\end{equation}
$\bar{a}_{\bf l\sigma}(\tau)= \exp(H_{\bf l}\tau)a^\dagger_{\bf
l\sigma}\exp(-H_{\bf l}\tau)$, and the first-order cumulant $K_1(\tau',\tau)=\langle{\cal
T}\bar{a}_\sigma (\tau')a_\sigma(\tau)\rangle_0$. This cumulant does not depend on the spin index $\sigma$ and the respective superscript is dropped in the cumulant notation. All operators in the cumulants belong to the same lattice site. Due to the translational symmetry of the problem the cumulants do not depend on the site index which is therefore omitted in the above equations. After the Fourier transformation the expression for $K_2^{\sigma'\sigma}$ reads
\begin{widetext}
\begin{eqnarray}
&&K_2^{\sigma'\sigma}(n_1,\nu+n_1,\nu+n_2,n_2)=
 Z^{-1}\Big\{\beta\Big[\big(\delta_{\nu 0}\delta_{\sigma'\sigma}
 -\delta_{n_1n_2}\big)
 e^{-E_1\beta}\nonumber\\
&&\quad+Z^{-1}\big(\delta_{\nu 0}-\delta_{n_1n_2}
 \delta_{\sigma'\sigma}\big)\Big(e^{-(E_0+E_2)\beta}-
 e^{-2E_1\beta}\Big)\Big]F(n_1+\nu)F(n_2)\nonumber\\
&&\quad-\delta_{\sigma',-\sigma}e^{-E_0\beta}Ug_{01}(n_1+\nu)
 g_{01}(n_2)g_{02}(n_1+n_2+\nu)
 \big[g_{01}(n_2+\nu)+g_{01}(n_1)\big]\nonumber\\
&&\quad-\delta_{\sigma',-\sigma}e^{-E_2\beta}Ug_{12}(n_1+\nu)
 g_{12}(n_2)g_{02}(n_1+n_2+\nu)
 \big[g_{12}(n_2+\nu)+g_{12}(n_1)\big]\nonumber\\
&&\quad+\delta_{\sigma',-\sigma}e^{-E_1\beta}\Big[F(n_1+\nu)
 g_{01}(n_2)g_{01}(n_2+\nu)+
 F(n_2)g_{01}(n_1+\nu)g_{01}(n_1)\nonumber\\
&&\quad+F(n_2)g_{12}(n_2+\nu)\big[g_{12}(n_1+\nu)-g_{01}(n_1)\big]+
 F(n_1+\nu)g_{12}(n_1)\big[g_{12}(n_2)-g_{01}(n_2+\nu)\big]\Big]\Big\},
\label{K2}
\end{eqnarray}
\end{widetext}
where $E_0=0$, $E_1=-\mu$, and $E_2=U-2\mu$ are the eigenenergies of
the site Hamiltonian~(\ref{siteH}), $\beta=T^{-1}$, $Z=e^{-E_0\beta}+
2e^{-E_1\beta}+e^{-E_2\beta}$ is the site partition function,
$g_{ij}(n)=(i\omega_n+E_i- E_j)^{-1}$, $F(n)=g_{01}(n)-g_{12}(n)$, and integers $n$ and $\nu$ stand for the fermion and boson Matsubara frequencies.

Equation~(\ref{K2}) can be significantly simplified for the case of
principal interest $U\gg T$. In this case, if $\mu$ satisfies the
condition
\begin{equation}\label{condition}
\eta<\mu<U-\eta,
\end{equation}
where $\eta\gg T$, the exponent $e^{-\beta E_1}$ is much larger than $e^{-\beta E_0}$ and $e^{-\beta E_2}$. Therefore terms with $e^{-\beta E_0}$ and $e^{-\beta E_2}$ can be omitted in Eq.~(\ref{K2}). In this case we obtain for the quantity $K_2=\frac{1}{2}\sum_{\sigma' \sigma}K_2^{\sigma'\sigma}$ which is used instead of $\lambda$ in Eq.~(\ref{Bethe2})
\begin{eqnarray}
&&K_2(n_1,\nu+n_1,\nu+n_2,n_2)\nonumber\\
&&\quad=-\frac{3}{4}\beta\delta_{n_1n_2}F(n_1+\nu)F(n_2)\nonumber\\
&&\quad+\frac{1}{2}\Big\{F(n_1+\nu)g_{01}(n_2)g_{01}(n_2+\nu)
 \nonumber\\
&&\quad+F(n_2)g_{01}(n_1+\nu)g_{01}(n_1)\nonumber\\
&&\quad+F(n_2)g_{12}(n_2+\nu)\big[g_{12}(n_1+\nu)-g_{01}(n_1)\big]
 \nonumber\\
&&\quad+F(n_1+\nu)g_{12}(n_1)\big[g_{12}(n_2)-g_{01}(n_2+\nu)\big]
 \Big\}.\quad \label{K2s}
\end{eqnarray}
As can be seen from Eq.~(\ref{Bethe2}), in this approximation $\Lambda({\bf k}_1,\omega_{n_1};{\bf k+k}_1,\omega_\nu+\omega_{n_1};{\bf k+k}_2,\omega_\nu+\omega_{n_2};{\bf k}_2,\omega_{n_2})$ does not depend on ${\bf k}_1$ and ${\bf k}_2$.

Taking into account these simplifications we find for the second term on the right-hand side of Eq.~(\ref{GF2})
\begin{eqnarray}
&&\left(\frac{T}{N}\right)^2\sum_{p_1p_2}\Lambda=-\frac{3}{4}T
 \sum_nf_{\bf k}(\nu n)a_1(\nu+n)a_1(n)\nonumber\\
&&\quad-\frac{1}{2}\big[1+S_{\bf k}y_4({\bf k}\nu)\big]T\sum_nf_{\bf
 k}(\nu n)a_3(-\nu,\nu+n)\nonumber\\
&&\quad+\frac{1}{2}\big[1-S_{\bf k}y_1({\bf k}\nu)\big]T\sum_nf_{\bf
 k}(\nu n)a_2(-\nu,\nu+n)\nonumber\\
&&\quad+\frac{1}{2}\bigg[\frac{1}{i\omega_\nu-U}-S_{\bf k}y_3({\bf
 k}\nu)\bigg]\nonumber\\
&&\quad\times T\sum_nf_{\bf k}(\nu n)a_4(-\nu,\nu+n)\nonumber\\
&&\quad-\frac{1}{2}S_{\bf k}y_2({\bf k}\nu)T\sum_nf_{\bf
 k}(\nu n)a_1(\nu+n),\label{scnd}
\end{eqnarray}
where $S_{\bf k}=N^{-1}\sum_{\bf k'}t_{\bf k+k'}t_{\bf k'}$,
\begin{eqnarray}
&&f_{\bf k}(\nu n)=\bigg[1-\frac{3}{4}S_{\bf k}F(\nu+n)F(n)\bigg]^{-1}, \nonumber\\
&&a_1(n)=F(n),\quad a_2(\nu n)=g_{01}(n)g_{01}(\nu+n), \nonumber\\[-1ex]
&&\label{terms}\\[-1ex]
&&a_3(\nu n)=F(n)g_{12}(\nu+n),\nonumber\\
&&a_4(\nu n)=g_{12}(n)-g_{01}(\nu+n),\nonumber
\end{eqnarray}
and $y_i({\bf k}\nu)$ are solutions of the following system of four linear equations:
\begin{eqnarray}
&&y_i({\bf k}\nu)\nonumber\\
&&\quad=T^2\sum_{n_1n_2}f_{\bf k}(\nu
 n_1)K_2(n_1,\nu+n_1,\nu+n_2,n_2)a_i(\nu n_1)\nonumber\\
&&\quad-\frac{1}{2}S_{\bf k}T\sum_n f_{\bf k}(\nu n)a_1(\nu+n)a_i(\nu
 n)y_2({\bf k}\nu)\nonumber\\
&&\quad-\frac{1}{2}S_{\bf k}T\sum_n f_{\bf k}(\nu
  n)a_2(-\nu,\nu+n)a_i(\nu n)y_1({\bf k}\nu)\nonumber\\
&&\quad-\frac{1}{2}S_{\bf k}T\sum_n f_{\bf k}(\nu
 n)a_4(-\nu,\nu+n)a_i(\nu n)y_3({\bf k}\nu)\nonumber\\
&&\quad-\frac{1}{2}S_{\bf k}T\sum_n f_{\bf k}(\nu
 n)a_3(-\nu,\nu+n)a_i(\nu n)y_4({\bf k}\nu).\label{syseq}
\end{eqnarray}
The system of equations~(\ref{syseq}) follows from Eq.~(\ref{Bethe2}),
$$y_i({\bf k}\nu)=T^2\sum_{n_1n_2}a_i(\nu n_1)\Lambda_{\bf k}(n_1,\nu+n_1,\nu+n_2,n_2).$$

In calculating the first term on the right-hand side of Eq.~(\ref{GF2}) we use the Hubbard-I approximation\cite{Hubbard} for the electron Green's function. Under condition~(\ref{condition}) this function reads
\begin{equation}\label{HubbardI}
G({\bf k}n)=\frac{i\omega_n+\mu-U/2}{(i\omega_n+\mu)(i\omega_n+\mu-U)
 -t_{\bf k}(i\omega_n+\mu-U/2)}.
\end{equation}

\section{Charge susceptibility}
\begin{figure}[t]
\centerline{\includegraphics[width=7.5cm]{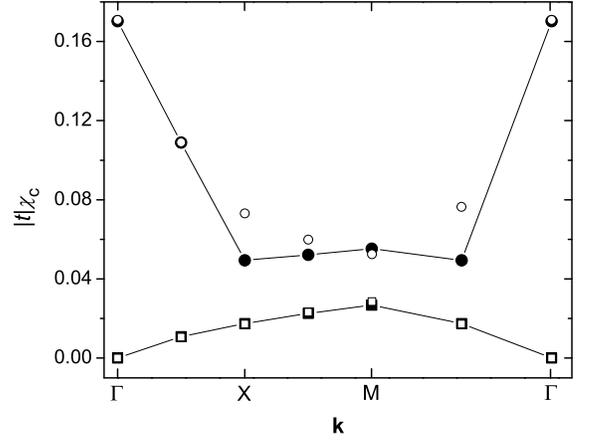}} \caption{The static charge susceptibility in a 4$\times$4 lattice for $t=-U/8$, $t'=0$, and $T=0.125|t|$. The susceptibility is plotted along the triangular contour in the Brillouin zone. The corners of the contour are given by the momenta ${\bf k}=(0,0)$ ($\Gamma$), $(\pi,0)$ ($X$), and $(\pi,\pi)$ ($M$). Open symbols are results of Monte Carlo simulations\protect\cite{Bickers} for $\bar{n}=1$ (squares) and $\bar{n}=0.95$ (circles). Filled symbols are our results for the same electron fillings.} \label{Fig2}
\end{figure}

In this section we consider the static charge susceptibility,
$$\chi_c({\bf k})=B({\bf k},\nu=0).$$
As mentioned above, divergence of this quantity points to charge instability of the system, while its finite maxima indicate regions of increased charge response on an external field. To check the used approximation we compare our results obtained for a 4$\times$4 lattice with the data of Monte Carlo simulations carried out for the same conditions.\cite{Bickers} This comparison is demonstrated in Fig.~\ref{Fig2}.
In this figure, $\bar{n}=\sum_\sigma\langle n_{\bf l\sigma}\rangle$ is the electron filling. The value $\bar{n}=1$ corresponds to half-filling. We used a small global vertical shift of our results to fit the Monte Carlo data better. As seen from the figure, our approximation reproduces correctly the general shape of the dependence $\chi_c({\bf k})$ and its variation with electron filling. Notice the rapid increase of the uniform susceptibility $\chi_c({\bf 0})$ with departure from half-filling.

\begin{figure*}
\centerline{\includegraphics[width=7.5cm]{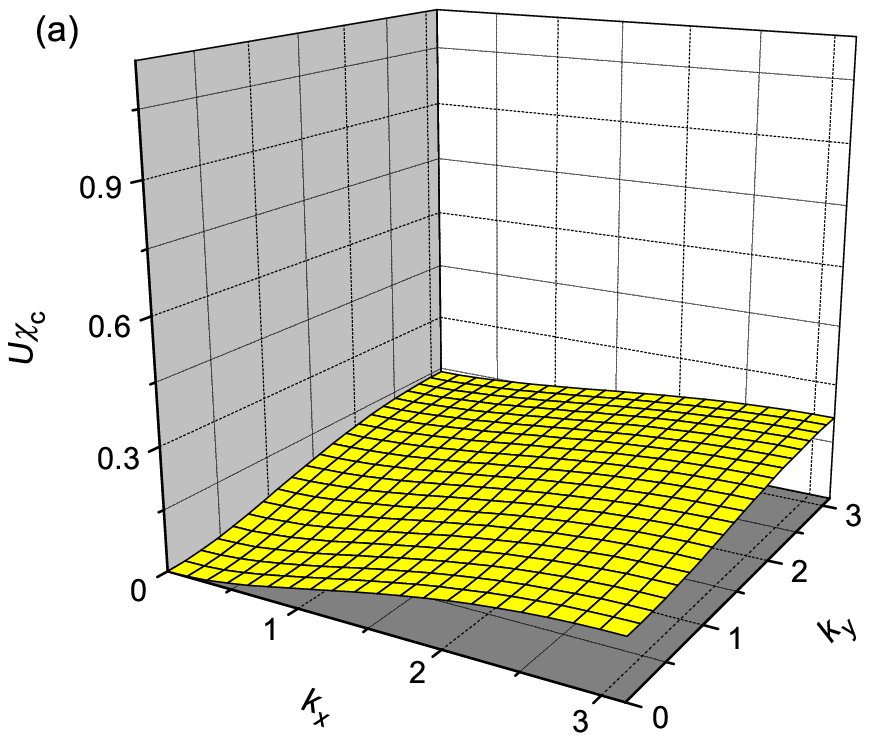}\hspace{3em}\includegraphics[width=7.5cm]{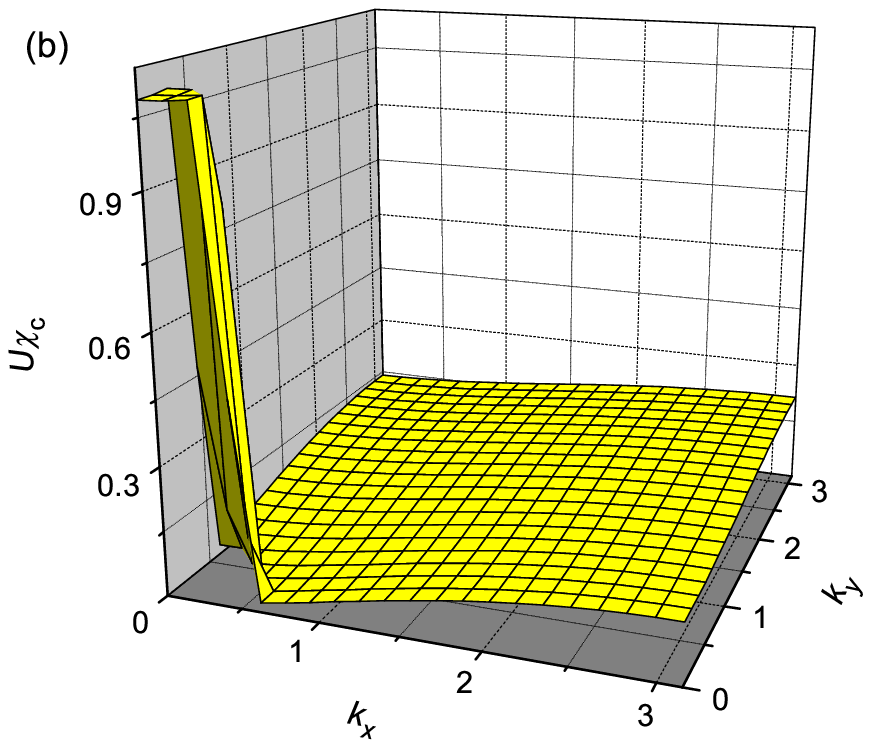}}

\vspace{3ex}
\centerline{\includegraphics[width=7.5cm]{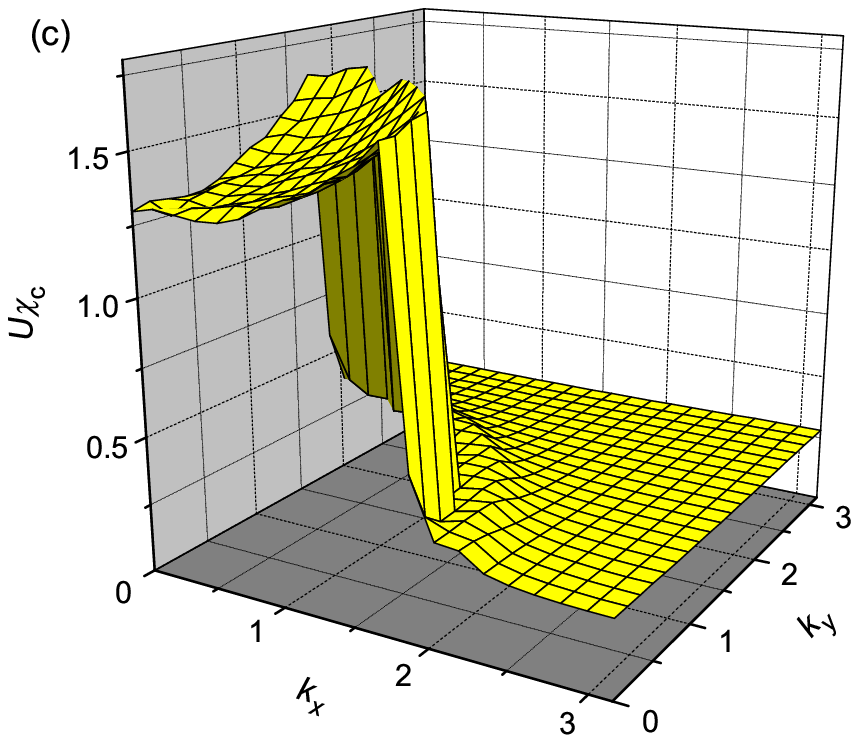}\hspace{3em}\includegraphics[width=7.5cm]{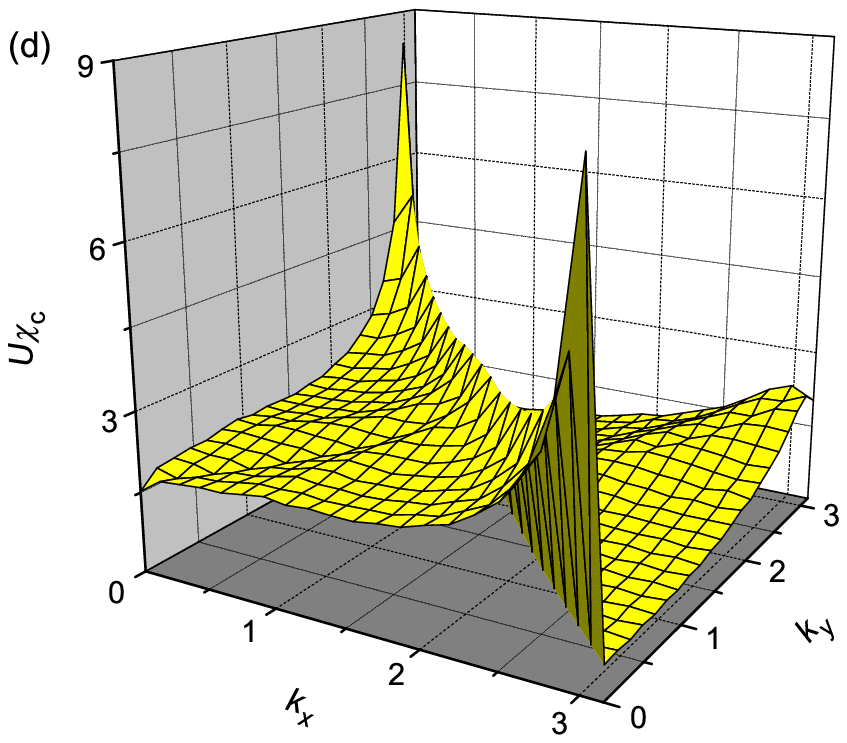}}
\caption{(Color online) The static charge susceptibility in a 40$\times$40 lattice for $t=-U/8$, $t'=0$, and $T=0.001U$. The susceptibility is shown in the first quadrant of the Brillouin zone for the electron fillings $\bar{n}=1$ (a), 0.97 (b), 0.88 (c), and 0.8 (d).} \label{Fig3}
\end{figure*}

The susceptibility for a larger lattice and a smaller temperature is shown in Fig.~\ref{Fig3}.
At half-filling the susceptibility is small and its dependence on momentum is weak, as can be seen in Fig.~\ref{Fig3}(a). This shape of $\chi_c({\bf k})$ is retained, until the Fermi level crosses one of the Hubbard subbands. Immediately after this crossing, which leads to departure from half-filling,  a sharp peak appears in the dependence $\chi_c({\bf k})$. As shown in Fig.~\ref{Fig3}(b), the peak is finite and is located near the $\Gamma$ point which indicates the appearance of long-wave charge fluctuation in the system. With increasing $x=|1-\bar{n}|$ the susceptibility grows. For moderate values of $x$ it peaks for momenta along a contour which is centered on the $\Gamma$ point and has the shape of a somewhat distorted circle [see Fig.~\ref{Fig3}(c)]. With increase of $x$ the size of the contour grows indicating the decrease of the wavelength of charge fluctuations. Notice that in Fig.~\ref{Fig3}(c), for $x \approx 0.12$, the contour crosses the axes at $k_x, k_y\approx\pi/2$. On approaching the boundary of the considered range of $\mu$, Eq.~(\ref{condition}), the susceptibility start to grow near the $X$ and $Y$ [$(0,\pi)$] points [see Fig.~\ref{Fig3}(d)]. For chemical potentials in this range the contour does not cross the boundary of the magnetic Brillouin zone -- the line segment connecting the $X$ and $Y$ points. Notice also that in this range corresponding to electron fillings $0\leq x\alt 0.2$ which are most interesting for cuprate perovskites the susceptibility remains finite. This means that a long-range or stripe ordering of charges {\em does not occur} in the Hubbard model for the considered parameters. The model produces only charge fluctuations with wave vectors corresponding to the maxima of the susceptibility. The susceptibility varies only slightly with a further decrease of temperature from the value used in Fig.~\ref{Fig3}.

\begin{figure}
\centerline{\includegraphics[width=7.5cm]{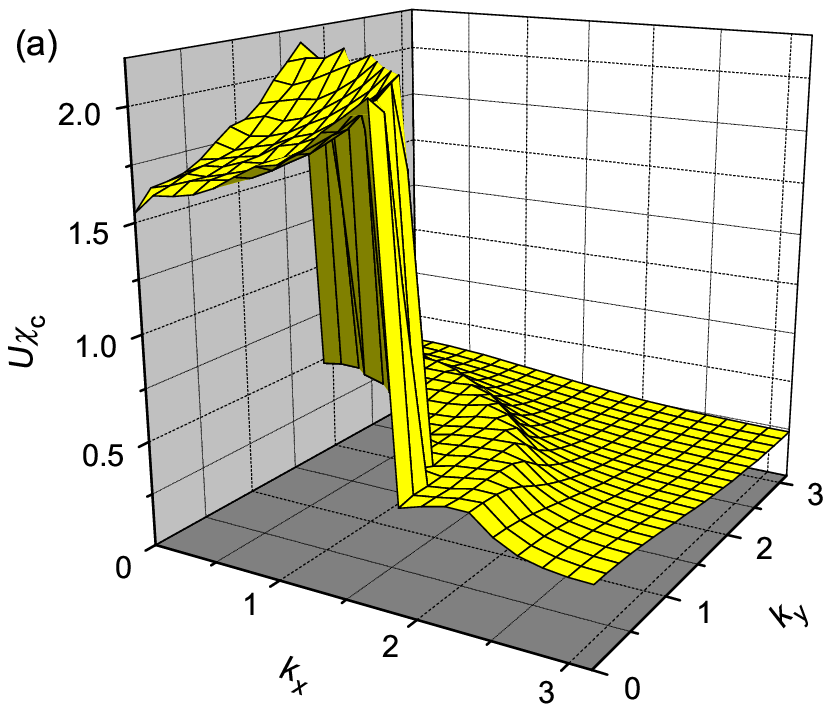}}

\vspace{4ex}
\centerline{\includegraphics[width=7.5cm]{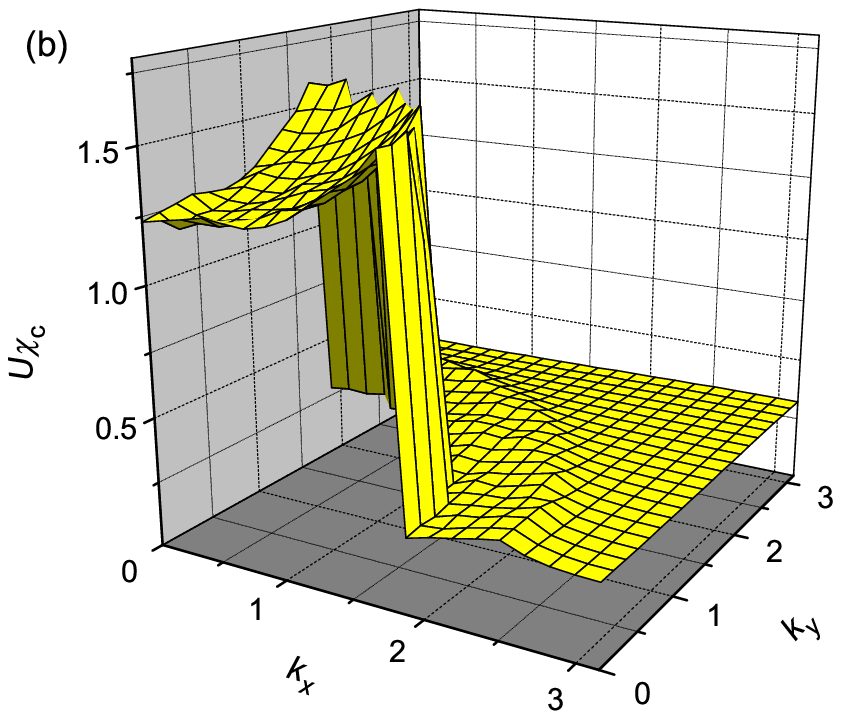}}
\caption{(Color online) The static charge susceptibility in a 40$\times$40 lattice for $t=-U/12$, $t'=0$ (a), and for $t=-U/8$, $t'=-0.3t$ (b). In both cases $\bar{n}\approx 0.88$ and $T=0.001U$.} \label{Fig4}
\end{figure}

The results shown in this figure were obtained for the $t$-$U$ Hubbard model with the ratio of parameters $U/|t|=8$. To check how this overall picture is changed with the variation of the parameters we carried out analogous calculations for the $t$-$U$ model with $U/|t|=12$ and for the $t$-$t'$-$U$ model with $U/|t|=8$ and $t'=-0.3t$. This latter ratio of the hopping constants is believed to describe adequately the band structure of cuprate perovskites. \cite{Korshunov} In both cases the evolution with filling and the shape of the susceptibility look qualitatively the same as those shown in Fig.~\ref{Fig3}. Moreover, the location of the susceptibility maxima in the Brillouin zone varies only slightly with the change of the hopping and repulsion constants and is mainly determined by the electron filling. As an example the susceptibilities for the two considered sets of parameters are shown in Fig.~\ref{Fig4} for $\bar{n}\approx 0.88$. These susceptibilities should be compared with Fig.~\ref{Fig3}(c). Notice that, as in Fig.~\ref{Fig3}(c), the maxima on the axes of the Brillouin zone are located approximately halfway between its center and boundary. It is worth noting also that in the $t$-$t'$-$U$ model, which does not possess the electron-hole symmetry, the shapes of the susceptibility are qualitatively similar for fillings $\bar{n}=1-x$ and $1+x$.

The appearance of maxima in the susceptibility with departure from half-filling is connected with fermion poles of the function $\Lambda$, specifically with the poles given by the function $f_{\bf k}(\nu n)$ in Eq.~(\ref{terms}). In general these poles do not coincide with poles of the electron Green's function $G(p)$. For $\nu=0$ we find four poles of the function $f_{\bf k}(\nu=0,z)$ at frequencies
\begin{equation}\label{poles}
\varepsilon_i({\bf k})=-\mu+\frac{U}{2}\pm\sqrt{\frac{U^2}{4}\pm \sqrt{\frac{3}{4}U^2S_{\bf k}}}.
\end{equation}
After the transformation to real frequencies sums over $n$ in Eqs.~(\ref{scnd}) and~(\ref{syseq}) are transformed to sums over the above poles and the terms of these sums contain multipliers $n_F(\varepsilon_i({\bf k}))$ where $n_F(\omega)=[\exp(\beta\omega)+1]^{-1}$ is the Fermi-Dirac distribution function. Thus, for low temperatures and small to moderate departure from half-filling these sums and the second term of $B(q)$, Eq.~(\ref{scnd}), will have a step-like shape as functions of momentum if one of the dispersions~(\ref{poles}) crosses the Fermi level. This behavior is demonstrated in Figs.~\ref{Fig3}(b), \ref{Fig3}(c), and~\ref{Fig4}.

The above results do not support a purely electronic mechanism of long-range charge ordering. However, they give some insight into the way in which phonons can stabilize stripes. As mentioned above, an essential role of certain CuO$_6$ octahedra tilts in such stabilization can be supposed from experimental results \cite{Tranquada95,Tranquada96,Fujita,Kimura} on lanthanum cuprates. It is known \cite{Pickett,Barisic} that such tilts are strongly coupled to the carriers. This interaction can be described using an effective electrostatic potential $\phi_{\bf l}$ which is connected with the charge fluctuations $\langle\delta n_{\bf l}\rangle$ arising in Cu-O planes as a consequence of the tilts. These fluctuations give the following contribution to the adiabatic potential:
\begin{equation}\label{apot}
\Delta E=-\frac{1}{2}\sum_{\bf l}\langle\delta n_{\bf l}\rangle \phi_{\bf l}=-\frac{1}{2}\sum_{\bf k}\chi_c({\bf k})\phi_{\bf k}^2.
\end{equation}
Here we took into account that the charge susceptibility is the response of the system on the external potential. In Eq.~(\ref{apot}) we allowed also for low frequency of the octahedra tilts.\cite{Pickett} The potential $\phi_{\bf k}$ depends on the coordinates $Q_{{\bf k}x}$, $Q_{{\bf k}y}$ of these vibrations. Due to the symmetry of the problem with respect to a reflection in a Cu-O plane this dependence contains only even powers of the coordinates and starts from quadratic terms,
\begin{equation}\label{epot}
\phi_{\bf k}=A_{\bf k}Q_{{\bf k}x}^2+B_{\bf k}Q_{{\bf k}y}^2+C_{\bf k}Q_{{\bf k}x}Q_{{\bf k}y}+\ldots
\end{equation}
Hence the contribution~(\ref{apot}) is a quartic function of the coordinates. As follows from Eqs.~(\ref{apot}) and~(\ref{epot}), finite values of the vibration coordinates give an energy gain and the larger the value of the susceptibility is, the larger energy gain can be achieved (the lattice stability is provided by other quartic and higher power terms of the adiabatic potential). For weak momentum dependencies of other parameters of the adiabatic potential the largest energy gain is achieved for momenta at which the susceptibility is peaked. From  symmetry considerations it can be supposed that for the LTT distortions ($Q_{{\bf k}x}\neq 0, Q_{{\bf k}y}=0$ and vice versa) this gain is achieved for maxima on the axes of the crystal plane, and for the LTO distortions ($Q_{{\bf k}x}=\pm Q_{{\bf k}y}$) on its diagonals. Based on the experimental results \cite{Tranquada95,Tranquada96,Fujita,Kimura} it is clear that for $x$ in the range $0.12\pm 0.02$ and for temperatures $T\alt 50$~K (the domain of the LTT phase \cite{Kumagai} in La$_{2-x}$Ba$_x$CuO$_4$) parameters of the adiabatic potential provide the lowest minimum for tetragonal distortions, and outside of this region for orthorhombic distortions. As can be seen in Figs.~\ref{Fig3}(c) and~\ref{Fig4}, for $x\approx 0.12$ the susceptibility maxima are located on the axes approximately halfway between the $\Gamma$ and $X$ ($Y$) points. Such maxima give the lowest energy for the CDW with the wavelength equal to four lattice spacings, as observed experimentally in the LTT phase of lanthanum cuprates. \cite{Tranquada95,Tranquada96,Fujita,Kimura} Such a CDW is commensurate with the lattice which is essential for its stability. On the contrary, for $x\approx 0.12$ an oblique CDW in the LTO phase is in general incommensurate [see Figs.~\ref{Fig3}(c) and~\ref{Fig4}] which can explain its more disordered character.\cite{Kimura}

Another conclusion which can be made from the analysis of experimental data \cite{Fujita,Kimura} and from the above results concerns the magnetic incommensurability observed \cite{Yoshizawa,Birgeneau} in $p$-type cuprate perovskites in a wide range of hole concentrations, namely that it is unlikely that the magnetic incommensurability is connected with charge stripes. In La$_{2-x}$Ba$_x$CuO$_4$ the LTT phase with stripes along Cu-O bonds exists in a narrow range of hole concentrations near $x=0.12$. \cite{Kumagai} Outside of this range, in LTO phase, stripes are more disordered \cite{Kimura} and oblique. Based on the assumption that magnetic incommensurability is connected with the stripes it is difficult to explain the location of the low-frequency incommensurate maxima in the magnetic susceptibility along the axes of the Brillouin zone in both phases and a smooth variation of the incommensurability parameter with doping through these phases. Mechanisms of the magnetic incommensurability in strongly correlated systems, which are not based on charge stripes, were considered in Refs.~\onlinecite{Tremblay,Sherman07,Sherman04,Eremin}.

\section{Concluding remarks}
In this paper we calculated the charge susceptibility of the two-dimensional repulsive $t$-$t'$-$U$ Hubbard model using the strong-coupling diagram technique. In this technique, on the assumption of strong electron correlations Green's functions are calculated using the expansion in powers of the hopping constant. The terms of this expansion are expressed by means of site cumulants of electron creation and annihilation operators. For a small lattice we found a good agreement of the results obtained in this approach with Monte Carlo simulations. In a larger lattice it was found that once the Fermi level crosses one of the Hubbard subbands a sharp peak appears in the momentum dependence of the static charge susceptibility near the $\Gamma$ point. With further departure from half-filling the peak transforms to a ridge around this point. These results were obtained for three sets of parameters: for the $t$-$U$ model with ratios of the parameters $U/|t|=8$ and 12, and for the $t$-$t'$-$U$ model with $U/|t|=8$ and $t'=-0.3t$. We found that the shapes of the susceptibility and the positions of its maxima are similar in all three cases and are mainly determined by the value of the electron filling $\bar{n}$. This suggests that the obtained shapes of the susceptibility are inherent in the parameter range of cuprate perovskites. In the considered interval of the electron filling $0\leq|1-\bar{n}|\alt 0.2$ the static susceptibility is finite which points to the absence of the long-range charge ordering. However, obtained results give some insight into the way in which phonons can stabilize stripes. For $|1-\bar{n}|\approx 0.12$ the susceptibility maxima are located halfway between the center and the boundaries of the Brillouin zone. In this case an interaction of carriers with the tetragonal distortions gives a maximal energy gain for the charge density wave which is oriented along the Cu-O bond and has the wavelength equal to four lattice spacings. Such stripes are observed in the low-temperature tetragonal phase of lanthanum cuprates. We supposed also that the obtained incommensurate location of the susceptibility maxima along diagonals of the Brillouin zone can be the reason of a more disordered character of oblique stripes in the low-temperature orthorhombic phase in comparison with commensurate stripes of the low-temperature tetragonal phase.

\begin{acknowledgments}
This work was partially supported by the ETF grant No.~6918.
\end{acknowledgments}

\end{document}